\documentclass{article}
\usepackage[square]{natbib} 
\usepackage[pdftex]{graphicx}
\pdfoutput=1

\usepackage[pdftex]{hyperref} 

\title{Restrictions on the dynamic growth of navigation accuracy in groups of animals} 

\author{Vladimir N. Binhi}

\date{}

\begin{document}
	
\maketitle   
Prokhorov General Physics Institute of the Russian Academy of Sciences; 
		38 Vavilov St., Moscow, 119991; 
		vnbin@mail.ru	\\	
		
\begin{abstract}

Many migratory animals regularly travel thousands of kilometers, exactly finding their destinations. It is assumed that migrants have both a compass sense to hold their course, and a map sense --- a kind of ``biological'' GPS --- to correct accumulated errors and determine right direction during navigation. Unlike the compass sense, the map sense is still not reliably explained. Therefore, an alternative way is discussed in literature to eliminate the errors --- the aggregation of animals in groups and their coordinated movement. Orientation accuracy of a group may be significantly higher because the errors caused by the action of a variety of casual factors are averaged. This idea, called the ``many wrongs principle,'' has been confirmed both in behavioral experiments and in the results of computer simulations. However, until now there was no analytical model that considered this effect and its limitations. Such a model is presented in this article. The model is given in terms of the course deviation angle and its variance. It includes a few parameters: individual sensitivity of the animal compass, the power of the herd instinct, and the level of random noise. These parameters determine, among other factors, the size of the group that is optimal for precise navigation. It is shown that there is a range of the noise level, at which  the aggregation of individuals into groups that do not exceed a certain size would be favorable. -- {PACS: 87.23.Cc}\\


\end{abstract}


\section{Introduction}

Migratory animals find their seasonal habitats surprisingly accurate, overcoming sometimes several thousand kilometers \citep {Galler.1972}. One possible explanation is associated with the fact that migrants have a compass sense (not necessarily magnetic) and can use it for orientation.

However, only the orientation ability, or a compass sense --- the choice of direction in the horizontal plane relative to the local benchmarks --- is not enough for finding a geographical destination \citep [e.g.,] [] {Wiltschko.and.Wiltschko.1995}. A variety of environmental factors --- the wind, geographical relief, the geographical and global perturbations of the magnetic field (MF), etc. --- can significantly influence the direction of movement.

It is widely believed, therefore, that in addition to the sense of direction, migrating animals must have the so-called map sense --- the ability to determine geographical position relative to the destination. The origin of this sense is not exactly known \citep {Gould.1982, Alerstam.2006, Kishkinev.ea.2015}. Cues of different nature may play a role: the sun, the stars, visual cues, the polarization of daylight, smells, weather, sounds, the geomagnetic field (GMF), and possibly others \citep {Rodrigo.2002}. Among all these cues, the GMF is the only global source of information that is sufficient for approximate positioning and independent of the weather.

The observed sensitivity of organisms to weak MFs is in general a fundamental physical problem \citep {Binhi.2002}, since the MF can change both the behavior of animals and the state of, e.g., cell cultures \citep{Belyaev.ea.2001}. However, the action of not only GMF variations but also three orders of magnitude higher MFs is not reliably explained, although there are a few candidate mechanisms. The basis of these are magnetic nanoparticles \citep [e.g.,] [] {Shaw.ea.2015}, spin-correlated biradicals \citep [e.g.,] [] {Dodson.ea.2013}, and a synergism of nanoparticles and biradicals \citep {Binhi.2006.BEM}. 
Less frequently discussed is the effect of weak MFs on the rotation of molecular groups \citep {Binhi.ea.2002.PRE}. 
These mechanisms are useful to explain the biological magnetic compass.

Below we consider the magnetic sense a basic one in orientation and navigation in animals, although the result is more general and is valid for the compass sense of any nature.

It is useful to clearly separate the two related concepts --- the ``map sense'' and ``navigation.'' 
While the map sense is the ability to determine a geographic position, navigation is broadly defined as the ability to reach a remote destination, or target, in a non-random way. Probably --- and this is the subject of further discussion --- the map sense is not absolutely necessary for navigation \citep [e.g.,] [] {Gould.2011}. Listed below are five arguments in favor of such a judgment.

First, simultaneous use of the compass and the map would not be parsimonious. In some cases, only one ability, the compass sense, seems to be enough. The compass, e.g., an inclination compass \citep {Wiltschko.ea.1972}, allows an animal 1) to be guided along the MF lines and 2) to select one of the two possible directions depending on whether the GMF vector projection on the longitudinal axis of the bird skull --- where the magnetic sensor is fixed --- decreases or increases with small trial pitches. This is a bicomponent compass. The navigational use of this compass is to follow certain tactics: to fly in the direction that is determined by the same local MF inclination, i.e., along the so-called isoclines \citep {Akesson.ea.2015}. A bird can also compare the value or inclination of the local MF with a preferred value recorded in its memory as a congenital or acquired cognitive map that is synchronized by the time of flight. This allows the bird to fly along the memory-specified course to familiar places, even in the absence of the map sense in its exact meaning, --- i.e., while not tracking its own geographical position relative to the target. It is only important that there is a preferred direction defined by the compass and memory in each local flight area --- the ``clock and compass'' mechanism. If birds adhere to the preferred directions, they approach their destination.

Second, map sense, or a biological GPS, --- the ability of an animal to determine its geographical location (two coordinates) and compare it with a cognitive standard --- requires at least two biosensors, one for each coordinate. In the case of magnetic biological GPS --- four sensors. Two reference vectors are necessary for measurement of the horizontal and vertical components of the geomagnetic field. One is given by the equilibrium sense --- it is the gravity vector. The second one must be supplied by some other independent source --- the polarization of daylight, direction to the sun, etc. The map sense thus is complicated. Probably, animals would have avoided its use in the presence of a simpler alternative.

Third, even in the case of using both the compass and the map, it seems fair to say that the use of the map in a continuous mode would not be optimal. For example, the traveler in the woods who is led by weak audio landmarks, is forced to walk long track sections ``blindly,'' while only occasionally stopping and listening, because motion worsens hearing markedly. Similarly, a bird can fly large parts of the migration path, guided only by the compass sense, and only occasionally change the mode of movement, which would allow it to activate the map sense.

Next, magnetic navigation strategies that are used by animals in migrations of different types --- approximately along the meridians and along the parallels --- are essentially different. In the first case, the GMF inclination is changing dramatically, and the use of the inclination compass might be enough. In the second case, preliminary construction of a map of the MF gradients seems necessary. \citet{Akesson.ea.2015} compared different compass mechanisms that could be allegedly involved in a long-distance migration. These were orientation by the stars, by the coastline, and by the GMF. They have concluded that following along the GMF isoclines was the most likely. This requires the ability to sense a) the local MF inclination, and b) the MF inclination gradient along the way, by flying some small parts of it. For the initial construction of the gradient map, which could be used for navigation later, a bird needs sufficiently long trial flights, controlled, of course, by the compass sense only.

Finally, many animals could probably take just approximately right course from the starting point, based on non-magnetic cues, although it is associated with some dispersal of the animals upon arrival at the destination. In addition, in cases where an inclination compass does not work --- in areas with a significantly reduced gradient of the GMF vertical component (magnetic equator) --- magnetic navigation ability is lost.

In all these cases, it is essential that animals can travel significant sections of the path accurately in the absence of the map sense. This is an orientational navigation. This raises the question of its performance under the action of concomitant interfering random factors.

It is evident that in the ideal case of no interference, target accuracy depends on the resolution of the compass sense or, for example, of a magnetic sensor. In turn, the resolution of the sensor is associated with its sensitivity to the MF. If the sensitivity is low, the animal is not able to navigate with respect to MF lines accurately; it can fail to meet a visually familiar landmark and will miss the target. The maximum sensitivity of the magnetic compass in birds, for example, is of the order of $\Delta H =  10$ nT \citep [e.g.,] [] {Kirschvink.and.Walker.1985, Ritz.ea.2009}. This gives angular resolution $ (\Delta H / H_ {\rm geo}) ^ {1/2} $ at the level of about $1^ {\rm o}$; an order lower resolution was also observed \citep {Gould.2011}. At this resolution, the random deviation from the target at the end of the migratory route of, for example, a thousand km length, would be $1$--$20$ km. It is acceptable, since the correction can be carried out by local visual, auditory, olfactory, and other cues.

However, many factors impede accurate navigational orientation. Among them are both external disturbances and physiological individual errors. As a result, the scattering of animals at the end of the migration path should significantly exceed an acceptable value, which is often in conflict with observations. One of the solutions to this problem was suggested half a century ago \citep {Bergman.and.Donner.1964} --- the direction of flock motion is the average of the individual directions, which improves the accuracy of the group migration. Since then, this idea, that is now called ``many wrongs principle'' \citep {Simons.2004}, received a qualitative consideration \citep {Walraff.1978} and several observational evidence. By comparing the parameters of many-kilometer flights of single pigeons and flocks of a few pigeons, \citet {Keeton.1970} has not found, while \citet {Tamm.1980} and \citet {DellAriccia.ea.2008} have found a difference in their orientation abilities. The growth of accuracy is evidenced on the basis of numerical simulations \citep [see] [and references therein] {Codling.ea.2007, Flack.ea.2015} and based on the characteristics of the collective motion of robots \citep {Gokce.ea.2010}.

The dependence of the accuracy of collective orientational navigation on the size of the group and limitations of this relationship have not yet been studied analytically.

The size of groups in animal aggregation is affected by various social factors --- foraging, breeding, predation pressure, etc. At migrations, the accuracy of navigation becomes one of the crucial factors. If the increase in the size of a group implies the growth of navigation accuracy, some migratory routes are only available for large groups, which has an evolutionary significance \citep {Walraff.1978}. In general, longer migration routes are available for larger groups. Such a correlation among the flocks of evolutionarily related species of birds is actually observed \citep {Beauchamp.2011}.

``Many wrongs'' is, in fact, a simple wording of the central limit theorem of mathematical statistics: many, $N$, errors mostly cancel each other out, so that the error of the mean for the group is reduced proportionate to $1 / \sqrt {N}$. However, this is not enough to determine the optimum size of the group. Below, we consider an analytical model of the orientational navigation and show that the individual tactics of following the group does lead to a significant increase in navigation accuracy. Important parameters of the model are the compass sensitivity, the power of herd instinct, and the level of external noise. The ratio of these parameters determines not only the accuracy of the collective orientational navigation, but also the size of group that is optimal for precise navigation.

For clarity, we speak about the navigation of birds, although the results, due to their generality, are applicable to all types of individuals provided they possess the compass sense and migrate collectively.

\section{A dynamic model of collective migration} \citet {Vicsek.ea.1995} have revealed collective effects in a system of particles with a biologically-motivated interaction. The system starts to behave as an independent entity to control the behavior of individual particles. In this model, an important parameter was the interaction radius: particles that are separated by a distance greater than a predetermined one do not interact. Collective orderly behavior was observed with an increase in the interaction radius or with an increase in the density of particles at a fixed radius.

In our model, the nature of the interaction may be connected both with vision and hearing.

Most birds have only monocular vision. Although there are several monocular mechanisms for the perception of the distance to an object, such as motion parallax and accommodation, we assume that a bird when migrating responds more to the visual density of the flock than allocates those birds that are near. Creating by the brain an averaged low-resolution picture in the wide field of view --- a mental building of the visual density --- can facilitate optimal orientation of an individual when moving in-group \citep {Wystrach.ea.2016}.

Also possible is the acoustic channel of interaction using the level of noise that is generated by the wings and voice signals. Close neighbors do not create main contribution to the noise level perceived by an individual, because the quadratic decrease of noise intensity with distance is compensated for by the growth in the number of individuals with distance, approximately also quadratic. Therefore, a bird can be oriented relative to the center of flock, where the noise is maximum.

Thus, an idealization of the infinite radius of interaction is justified for both the visual and acoustic channel. \citet {Vicsek.ea.1995} demonstrated that in this case, the collective effects are maximized.

As mentioned above, a bird can migrate along a certain trajectory even in the absence of a map sense, if in each local region there is a local cue --- a preferable direction that is defined by the compass sense and memory, or a cognitive map. This cue can slowly change over time as the bird moves forward. Use of this cue ensures the achievement of destination even if the flight path is not straight. This fact allows one to consider a mathematical problem as essentially a one-dimensional problem, in spite of the two-dimensional character of the motion. The problem is defined in terms of the angle of the deviation from the local preferable direction.

Other idealization used to build the model are as follows. a) A bird has the compass sense; the orientation of the flock is realized only through the compass sense of individuals. b) An individual knows, where the flock is, and tries to keep the same direction of flight with the flock.  c) All individuals are equal, and there are no pair interactions, in particular collisions --- this enables a simple analytical study. d) The orientation of the flock-as-a-whole is identified with the average orientation of individuals, i.e., the ``center'' of the flock in terms of angles.

A stochastic equation for the orientation of a single $i$-th individual has the form \begin{equation} \label{eqspec} \frac{{\rm d} \varphi _i }{{\rm d} t}  = -m\varphi _i - k \left( \varphi _i - \alpha \right) + \xi _i  \end{equation} where $\varphi _i$ is the individual angle of deviation from the local preferable direction to the target, or destination, $ \alpha \equiv (1/N) \sum_i \varphi _i $ is the angle of the deviation of the flock from the preferable direction to the target, Fig.\,\ref{IndFlock}, $ t $ is time, $ N $ is the number of individuals in the flock, $m$ and $k$ are constants, $ \xi _i $ is a random perturbation of power $w$. Angles are given in radians. We assume that $ \xi _i $ are the independent realizations of the white noise process with a limited bandwidth: $ \langle \xi_i \xi_j \rangle = 0, i \neq j $ and $\langle \xi_i (t) \xi_i (t-\tau ) \rangle = w \delta (\tau )$, where the angle brackets denote averaging over the ensemble of realizations, $ \delta (\tau) $ --- the Dirac delta function (with the dimension $ t ^ {-1}$). The first term of the equation describes the compass sense of an individual, and the second one --- the tendency of the individual to equate its direction of flight with the direction of the flock's flight. The real compass sensor with a certain angular resolution is presented as the ideal compass (first term) and a noise that is included in $ \xi $.

\begin{figure}[htbp] \def\figuretitle{IndFlock} \centering \includegraphics[scale=0.48]{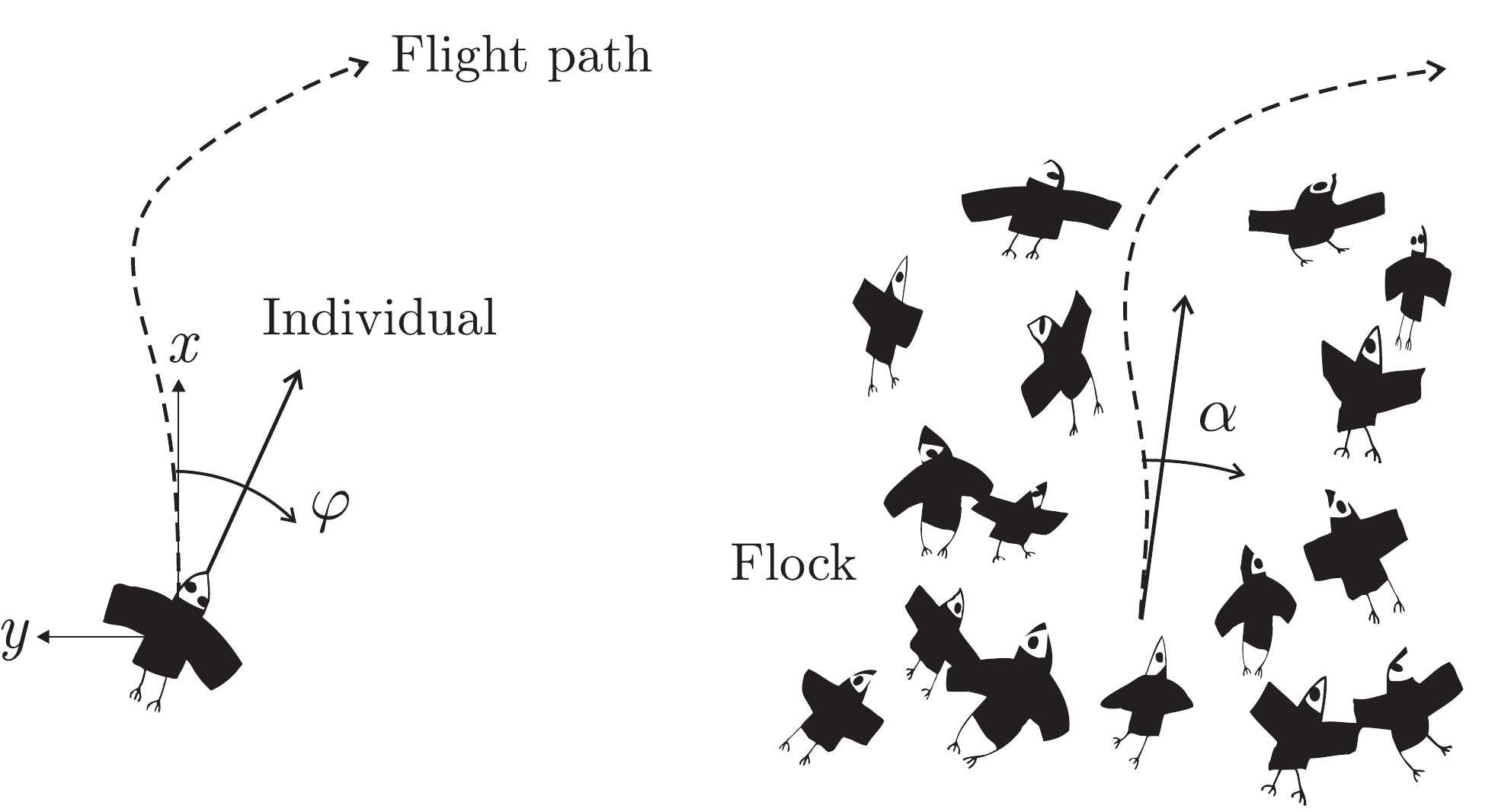} \caption{Directions of the flight of an individual and the flock relative to the local preferable direction $x$. }
\label{\figuretitle} \end{figure}

Note that the cyclical nature of variable $\varphi $ cancels a logarithmic response to the deviations from the course and the use of the Weber-Fechner law with regard to the animal perception in this case. Therefore, reactions to the deviations are written in the equation (\ref {eqspec}) in the linear approximation. In addition, the range of changes in $ \varphi $ that is expected in the process of navigation is small, well within $(- \pi / 2, \pi / 2)$. Larger deviations would mean that the bird is flying perpendicular to, or even against the movement of the flock. Although, in general, hairpin turns are natural and observable in a flock's flight, it is unlikely that such a motion is the part of the compass sense implementation during the process of a stationary migration.

Acting on equation (\ref{eqspec}) by operator $(1/N)\sum _i$, we find the equation of the flock's motion: \begin{equation}\label{eqflock}  \frac{{\rm d} \alpha }{{\rm d} t} = -m\alpha + \eta, ~~\eta\equiv \frac {1} {N} \sum_i \xi _i \end{equation}

An important characteristic of the solution to stochastic equation (\ref{eqflock}) is its variance, showing how far can the flock deviate from the course in average, while implementing its orientational navigation ability. In what follows, we find a variance $ \langle \alpha ^ 2 (t) \rangle $ at $ t \rightarrow \infty $, and compare it with the variance of the deviation of an individual flight in the flock. The variance can be written in this form, since the average value of $ \langle \alpha (t) \rangle $ is zero by virtue of the symmetry of the problem.

A solution to equation (\ref {eqflock}) with the initial condition $\alpha (0) = 0 $ without loss of generality is
\begin{equation} \label{asol} \alpha (t) = {\rm e}^{-mt} \int_0^t \eta (t) \,{\rm e}^{mt} {\rm d} t \end{equation} 
We write, then, using Fubini's theorem, the variance of angle $ \alpha $ in the following way: 
\begin{eqnarray}\label{asolb} \langle \alpha ^2 \rangle = \Big\langle {\rm e}^{-2mt} \int_0^t \eta (u) \,{\rm e}^{mu} {\rm d} u   \int_0^t \eta (v)\, {\rm e}^{mv} {\rm d} v   \Big\rangle \\ =  {\rm e}^{-2mt} \int_0^t\!\!\!\int_0^t \langle \eta (u) \eta (v) \rangle \, {\rm e}^{m(u+v)} {\rm d} v  {\rm d} u \nonumber \end{eqnarray} 
From the definitions of $\eta $ and $\xi $ follows that $ \langle \eta (u) \eta (v) \rangle = (w / N) \delta (u-v) $, wherefrom we find 
\begin{equation}\label{adisp} \langle \alpha ^2 \rangle = \frac {w}{2mN}\left( 1- {\rm e}^{-2mt}\right) \end{equation} 
It can be seen that the variance of the deflection angle of a flock is inversely proportional to the number of individuals in the flock. Accordingly, the standard deviation $ \sigma _ {\alpha} \equiv \sqrt {\langle \alpha ^ 2 \rangle} $, i.e., the error of the mean of $\varphi_i$, decreases in proportion to $ 1 / \sqrt {N} $, which is in line with the ``many wrongs'' hypothesis. However, this figure does not give a correct idea of the navigational accuracy, because the ``center'' of flock can accurately hit the target, while the individuals are essentially dispersed and are far from the target. Then one cannot speak about precise navigation. Interest, thus, is in variance of the deflection angle of an individual, provided the flock has a controlling influence on it. We calculate this value.

Comparing equations (\ref {eqspec}) and (\ref {eqflock}), one can note that the first one is mathematically equivalent to the second one. Therefore, instead of a solution to equation (\ref {eqspec}), one can immediately write its variance in a form similar to (\ref {asolb}), however making in this expression replacements $ m \rightarrow m + k $ and $ \eta \rightarrow k \alpha + \xi_i $: \begin{equation}\label{fisol} \langle \varphi_i ^2 \rangle = {\rm e}^{-2(m+k)t} \int_0^t \!\!\!\int_0^t \big\langle [k\alpha (u) +\xi_i]  [k\alpha (v) +\xi_i]  \big\rangle \, {\rm e}^{(m+k)(u+v)} {\rm d} v  {\rm d} u \end{equation} We introduce the following notation: $x_1 \equiv \langle \alpha (u) \alpha (v) \rangle$, $x_2 \equiv \langle \alpha (u) \xi_i (v) \rangle $, $x_3 \equiv \langle \xi_i (u) \xi_i (v) \rangle $. Then, omitting index $i$ that is no longer necessary, we find \begin{equation} \label{fivar} \langle \varphi ^2 \rangle = {\rm e}^{-2(m+k)t} \big( k^2 I_1 + 2 k I_2 + I_3 \big) \end{equation} where \begin{equation} \label{manyI} I_{1,2,3} \equiv \int_0^t\!\!\!\int_0^t x_{1,2,3} \,{\rm e}^{(m+k)(u+v)} {\rm d} v  {\rm d} u \end{equation} To transform (\ref {fivar}), it needs to find the values of functions $x$ in (\ref{manyI}). From equations (\ref {eqflock}) and (\ref {asol}) it follows that \begin{equation}\label{xs} x_1  = \frac{w}{mN} \left\{  \begin{array}{ll} {\rm e}^{-mu} \sinh (mv), & u \geq v \\ {\rm e}^{-mv} \sinh (mu), & u<v \end{array}  \right. , ~~ x_2  = \frac{w}{N} \, {\rm e}^{m(u-v)} , ~~  x_3 = w \delta (u-v ) . \end{equation} Calculation from (\ref{fivar}--\ref{xs}) of function $ \langle \varphi ^ 2 \rangle $ of variables $m$, $k$, $N$, $w$, and $t$ leads to expression
\begin{eqnarray} \langle \varphi ^2 \rangle \frac {2N}{w} =  \frac{1}{m} \frac{k+2m+m n}{k+m} -\frac{4}{k+2m} {\rm e}^{-kt}  \\+ \frac{3k+2m- n(k+2m)}{k^2 +3km +2m^2} {\rm e}^{-2(k+m)t} - \frac{1}{m} {\rm e}^{-2mt} \nonumber \end{eqnarray}

\vspace{3mm} \noindent Functions $ \langle \varphi ^ 2 \rangle $ and $ \langle \alpha ^ 2 \rangle $ are shown in Fig.\,\ref {VarTime}. It can be seen that the variance of the deviation angles increases at first and then reaches saturation. The variance of the deviation of flock is significantly less than that of an individual.

\begin{figure}[htb] \def\figuretitle{VarTime} \centering \includegraphics[scale=0.38]{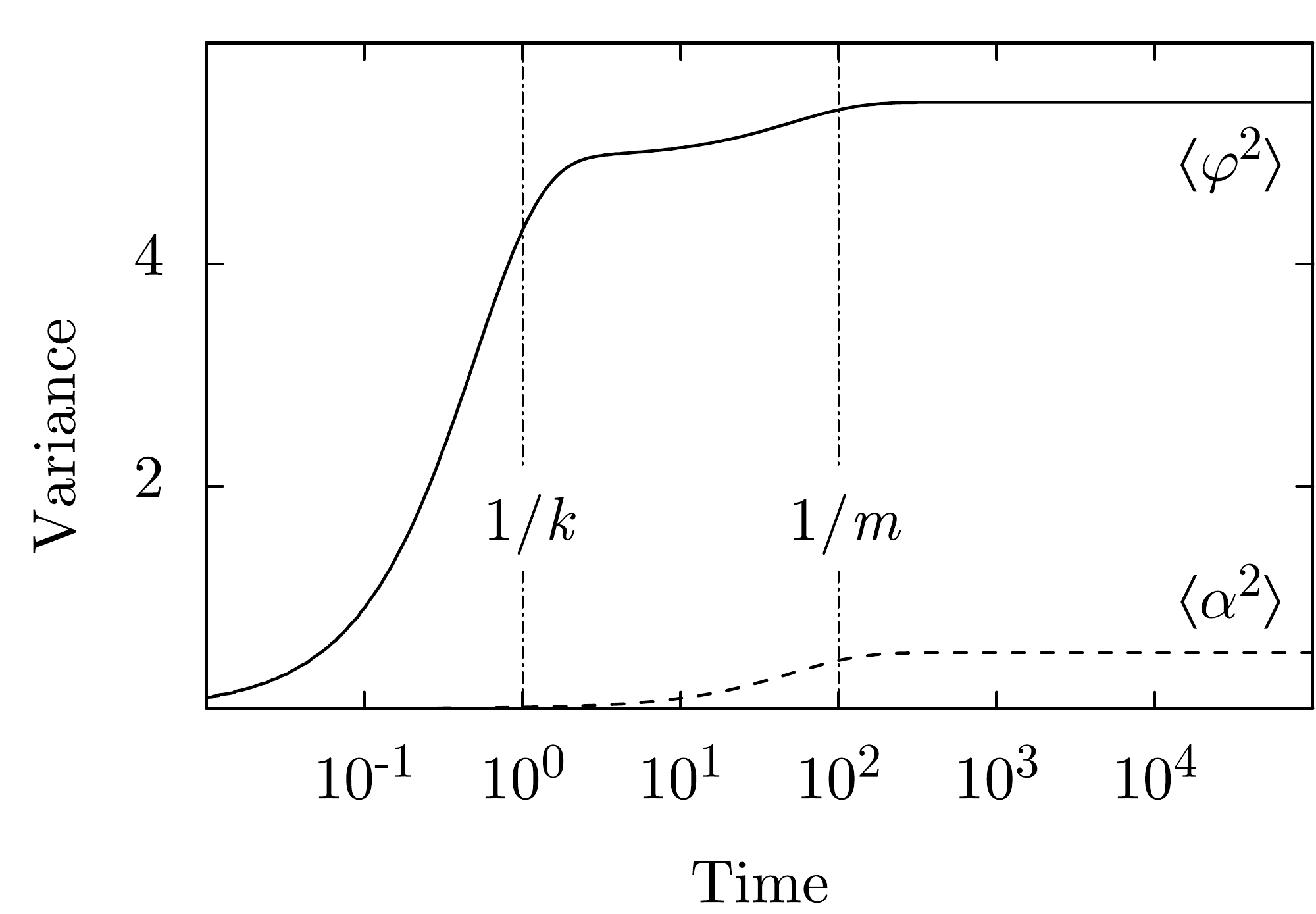} \caption{Time dependence of the deflection angle variances of a bird and the flock; $k=0.1$, $m=0.001$, $N=1000$, $w=1$.} \label{\figuretitle} \end{figure}

Reasonable value of the time constant, $1/m$, of the compass sense is clearly less than a few minutes (and much more so regarding $1/k$), while the time of migration flights is of several hours or more. Therefore, evaluation of the accuracy of navigation in the limit $t \gg 1/m$ is completely justified. In this limit, the variances are as follows: \begin{equation} \label{fatlarget} \langle \alpha ^2 \rangle \rightarrow \frac {w}{2mN}, ~~ \langle \varphi ^2 \rangle \equiv \sigma ^2  \rightarrow \frac{w}{2mN} \frac{k + 2m + m N}{k+m} \end{equation} where $ \sigma $ is a standard deviation. Functions (\ref {fatlarget}) are shown in Fig.\,\ref {VarNumber} depending on the number $N$ of individuals in the flock. If $N$ is large enough and exceeds $k/m$, the variance of the deflection angle of an individual is almost unchanged, \begin{equation} \label{fionN} \langle \varphi ^2 \rangle \approx w/2k , \end{equation} while the variance of the deflection angle of the flock continues to decrease in proportion to  $1/N$.

\begin{figure}[htb] \def\figuretitle{VarNumber} \centering \includegraphics[scale=0.41]{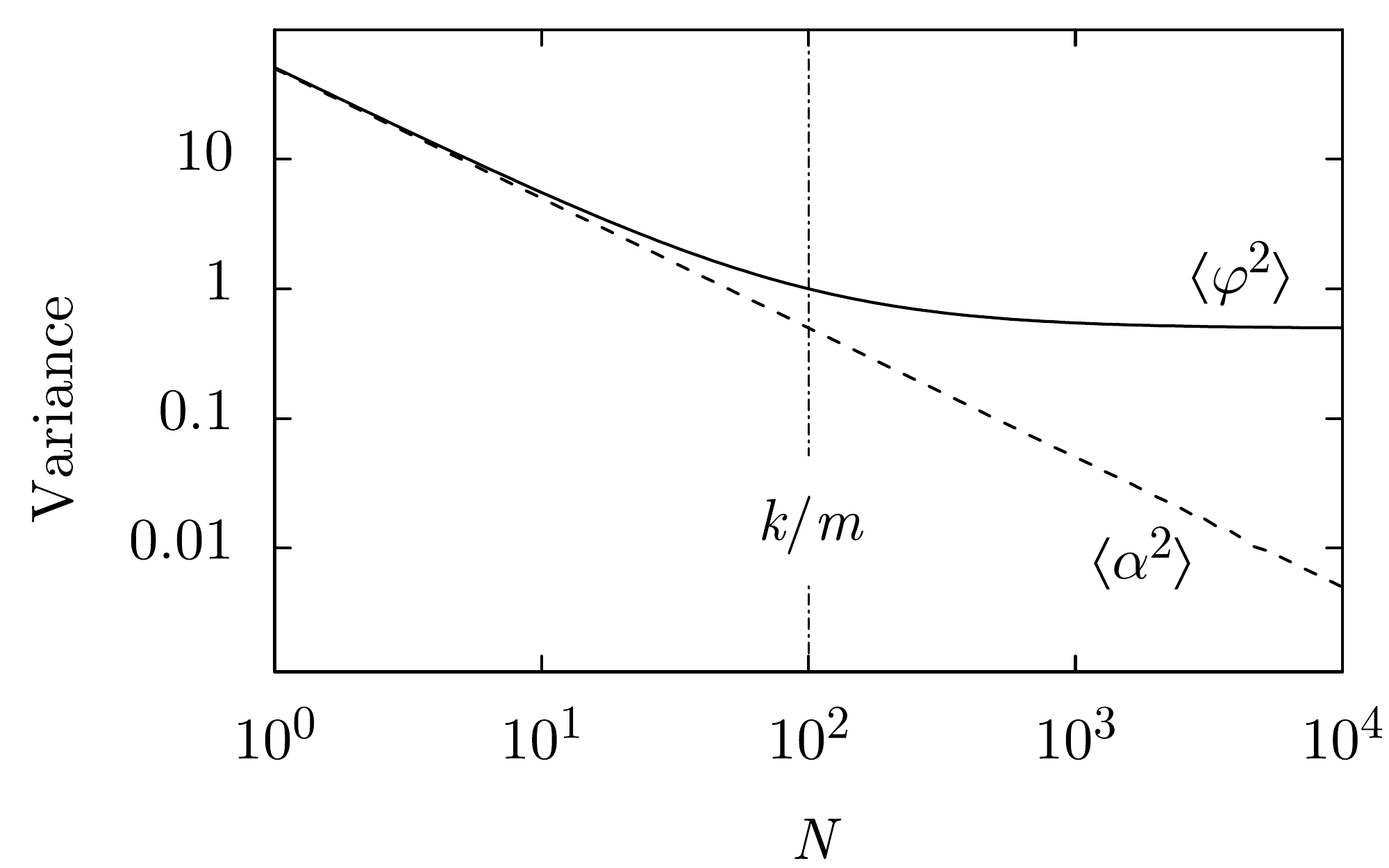} \caption{The dependence of the deflection angle variances of a bird and the flock on the number of individuals at $t\gg 1/m$; $k=1$, $m=0.01$, $w=1$.} \label{\figuretitle} \end{figure}

There are certain biological restrictions on the values of the parameters that control the orientation in the flock. It is natural to think that the implementation of the individual compass sense is a process that is slower than that of correcting the direction of individual flight in accordance with the direction of the flock. Mathematically, this means $m \ll k$. In other words, in migration, the compass sense is weaker than the herd instinct. The sensitivity of the magnetic compass in some animals is approximately $10$ nT. Biological integration of magnetic signals with such precision may require about $3.5$ s  \citep {Kirschvink.and.Walker.1985}; then, in this case, $ m \sim 0.3 $ s$^{-1}$. On the other hand, it takes a bird less than a second to determine its position in the flock, make a decision, and change the direction of flight \citep [e.g.,] [] {Nagy.ea.2010}. Our estimates of the characteristic constants are $ k \sim 1$--$10$ s$^{-1}$ and $ m \sim 0.01$--$1$ s$^{-1}$; these determine the optimal range of $N \sim k/m$ on the order of $1$--$1000$.

\section{Discussion} Despite the fact that the ``many wrongs'' hypothesis correctly reflects the increase in the accuracy of hitting the target by the flock represented by its center, it does not reflect the real navigation accuracy. The latter is determined by the accuracy of hitting the target by single individuals, i.e., by the standard deviation $\sigma $, see (\ref{fatlarget}).

At large times, the navigation accuracy with increasing $N$ at first increases. The dispersal of individuals decreases in proportion to $1/ \sqrt{N}$ that corresponds to the ``many wrongs'' hypothesis. However, with the further growth in $N$, the effect of increasing accuracy weakens and then virtually stops at $N \sim k/m$, Fig.\,\ref {VarNumber}. An even greater increase in the flock size has no biological meaning, because the navigation performance does not improve. Probably this, among other factors, determines the number of individuals in a migrating flock. Of course, if the size of the target area is great and larger than $L \sigma $, where $L$ is the length of the migration path, precise navigation can be achieved at lower values of $N$. We examine this in more detail.

One should make here a reservation that the model works in a 1D angular space and does not take into account the relative space positions of birds and their ability to adjust these positions. Such a regulation only partially converts to the control of orientations and so affects the dispersal of animals. However, taking space positions into account while improving the model would make it a two-dimensional one at least, and an analytical approach would be impossible.

Fig.\,\ref{IndFlock} implies that the differential of the individual deviation from the preferable direction is ${\rm d} y = v \sin {[\varphi (t)]} {\rm d} t$, where $v$ is the flight speed that is assumed to be constant. Integrating along the flight path, we get, in the end of the path, deviation $Y = v \int_0 ^ T \sin {[\varphi (t)]}\, {\rm d} t$, where $T = L/v $ is the travel time. $Y$, just like $\varphi $, is a random variable. It can be shown that in the approximation of small angles, the standard deviations of these variables are linked by equation $ \sigma_Y \approx L \sigma $.

The probability $p$ that a flock of $N$ birds will achieve a small-sized target area is directly proportional to its size $b$ and inversely proportional to the standard deviation $ \sigma_Y$, if $b < \sigma_Y$. Otherwise, obviously, $p = 1$. Since $b / \sigma_Y = b/L \sigma $, then \[  p(N) = \left\{ \begin{array}{rr} \beta / \sigma, & \beta / \sigma <1 \\ 1, & \beta /\sigma \geq 1 \end{array}  \right. \] where the notation $ \beta \equiv b / L $ is introduced --- it is the angular size of the target area that would be visible from the starting point, if the trajectory of the flight was straightened to a right line. From (\ref {fatlarget}), we find \[ \frac{\beta } {\sigma } = \frac {\beta }{\sqrt{w}}  \left(  \frac{{2mN} (k+m) } {k + 2m + m N}\right)^{1/2}   \]

The probability $p (N)$ is shown in Fig.\,\ref {ProbNumber} at different values of parameter $ \beta / \sqrt {w} $. It can be seen that for small random disturbances $w$ or for large target sizes $\beta $, a group of even several individuals reaches the target. At very small disturbances, the target is achievable even in the flight of a sole bird. On the contrary, at large disturbances, the target cannot be reached with a probability $p =  1$ by any arbitrarily large flock, and probability of achieving the target is $\beta \sqrt {2k / w}$.

\begin{figure}[htbp] \def\figuretitle{ProbNumber} \centering \includegraphics[scale=0.4]{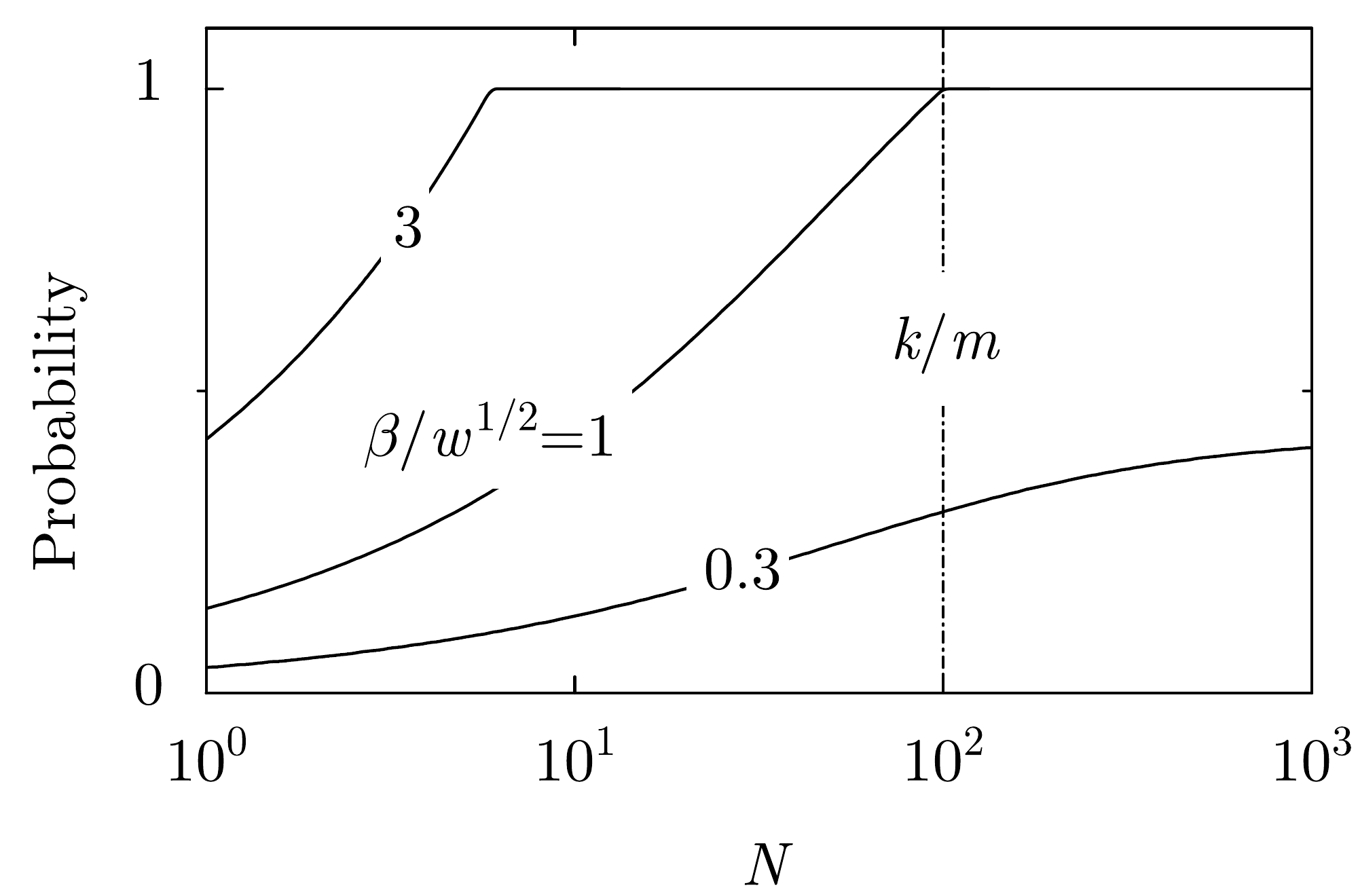} \caption{Probability of achieving the target area of size $\beta L$ by a group of size $N$. Shown are values of $\beta / \sqrt {w}$ from $0.3$ to  $3$; $k = 1$, $m = 0.01$. } \label{\figuretitle} \end{figure}

There is thus a certain range of random disturbances, in which the  aggregation of individuals in groups of no more than a certain size is beneficial. This is generally consistent with observations of groups of people who performed orientation tests \citep{Faria.ea.2009}.

Accuracy of navigation at optimal values $N \sim k/m$ is virtually independent of the responsiveness $m$ of the biological compass. The accuracy is defined in terms of standard deviation $\sigma \approx \sqrt {w/2k}$ depending only on the level of random disturbances $w$ and the rate $k$ of bird reaction to the change in the direction of flock's flight. In this case, the standard deviation $ \sigma $ at the end of the path does not exceed $\pi /2$ --- this latter value would mean an absolute scattering. Then, from (\ref{fionN}) we get $w < \pi ^ 2 k/2$. Since parameter $k$ reflects a genotypic characteristic that ensures the existence of individuals in flock and the flock itself, then this inequality imposes a limit on the power of random disturbances $w$, at which a flock can fly as a whole entity, and hence migration is possible. If so, a migration cannot begin until 1) the number of individuals at the start area is sufficient and 2) random disturbances are small enough, e.g., in windless weather. For example in \citep {Bergman.and.Donner.1964}, highest migration intensity was observed at a weak tailwind.

For small \textit {external} disturbances, noise power $w$ is determined by the resolution of the compass sensor. In this case, the formation of flocks for migration can compensate for the low sensitivity of the compass sensor. For example, some species of birds could be capable of precise collective navigation in favorable conditions even with a weak compass sense.

The above-considered increase in the accuracy of orientational navigation is a collective and \textit {dynamic} effect. It can only be realized with continued purposeful movement of a group. This mechanism is not relevant, for example, to the accuracy of individual primary (without a test flight) orientation to the target \citep [e.g.,] [] {Kishkinev.ea.2013}. This orientation is of compass rather than navigational nature.

This navigation model connects the optimal size of a group with the average individual characteristics and the noise level. At the same time, it is clear that the size of a real flock is also controlled by other factors, such as foraging, protection from predators and competitors, and social relations, which may acquire priority in the different phases of migration. Therefore, our numerical estimates are approximate. However, they correctly reflect the qualitative laws that limit the size of an orientationally navigating flock in the case of decrease in the importance of other factors.

Stochastic equation of orientation of an individual (\ref{eqspec}) can be split into two equations separately for the dominant leader and subordinate followers. This allows one to identify patterns of social effects in biological navigation that are intensively studied now \citep {Nagy.ea.2010, Codling.ea.2016}. Derivation of analytical relations in this case is cumbersome and will be the matter of a separate study.

\section{Conclusions} Apparently, migrating animals are capable of cognitively averaging a visual or audible pattern with the purpose of orienting themselves in the same direction as the group-as-a-whole velocity. This ability transforms into the strengthening of the group compass sense, or the improvement of navigational performance. Simple individual tactics --- to follow the flock --- significantly improves the chances of solving the strategic objective to achieve a seasonal residence. This effect is obviously associated with the central limit theorem, according to which the relative level of the sample mean fluctuations decreases with increasing the sample size as $1 / \sqrt {N}$. With respect to the ``many wrongs'' hypothesis, the effect, however, is limited. Very large values of $N$ are useless in terms of navigation accuracy. The responsiveness of the biological compass and the power of the herd instinct ($m$ and $k$) determine, among other possible factors, a threshold in the number of individuals, beginning from which a further growth of the group does not lead to an advantage in biological navigation. The nature of this limitation stems from the fact that the characteristics of a large flock cannot be fully represented by the characteristics of its center. The range of deviations of an individual that is guided by the flock becomes a key factor of navigation accuracy.

The presented dynamic model of collective migration describes the main features observed in the impact of animal aggregation on the orientational navigation. These are the growth of accuracy with an increase in the size of group to a certain optimum level and the absence of this effect in the case where random disturbances are small. The result is useful for the study of collective behavior and navigation.

\section*{Acknowledgements}

The author is grateful to D. A. Kishkinev for useful comments on the earlier version of this text.


\begin{thebibliography}{}
	
	\bibitem[Akesson and Bianco, 2015]{Akesson.ea.2015}
	Akesson, S. and Bianco, G. (2015).
	\newblock Assessing vector navigation in long-distance migrating birds.
	\newblock {\em Behavioral Ecology}, Published online, 31 December:1--11.
	
	\bibitem[Alerstam, 2006]{Alerstam.2006}
	Alerstam, T. (2006).
	\newblock Conflicting evidence about long-distance animal navigation.
	\newblock {\em Science}, 313(5788):791--794.
	
	\bibitem[Beauchamp, 2011]{Beauchamp.2011}
	Beauchamp, G. (2011).
	\newblock {Long-distance migrating species of birds travel in larger groups}.
	\newblock {\em Biology Letters}, 7(5):692--694.
	
	\bibitem[Belyaev and Alipov, 2001]{Belyaev.ea.2001}
	Belyaev, I.~Y. and Alipov, E.~D. (2001).
	\newblock {Frequency-dependent effects of ELF magnetic field on chromatin
		conformation in \textit{Escherichia coli} cells and human lymphocytes}.
	\newblock {\em Biochimica et Biophysica Acta}, 1526(3):269--276.
	
	\bibitem[Bergman and Donner, 1964]{Bergman.and.Donner.1964}
	Bergman, G. and Donner, K.~O. (1964).
	\newblock {An analysis of the spring migration of the Common Scoter and the
		Long-tailed Duck in southern Finland}.
	\newblock {\em Acta Zoologica Fennica}, 105:1--59.
	
	\bibitem[Binhi, 2002]{Binhi.2002}
	Binhi, V.~N. (2002).
	\newblock {\em {Magnetobiology: Underlying Physical Problems}}.
	\newblock Academic Press, San Diego.
	
	\bibitem[Binhi, 2006]{Binhi.2006.BEM}
	Binhi, V.~N. (2006).
	\newblock Stochastic dynamics of magnetosomes and a mechanism of biological
	orientation in the geomagnetic field.
	\newblock {\em Bioelectromagnetics}, 27(1):58--63.
	
	\bibitem[Binhi and Savin, 2002]{Binhi.ea.2002.PRE}
	Binhi, V.~N. and Savin, A.~V. (2002).
	\newblock Molecular gyroscopes and biological effects of weak extremely
	low-frequency magnetic fields.
	\newblock {\em {Physical Review E}}, 65(5):051912.
	
	\bibitem[Codling and Bode, 2016]{Codling.ea.2016}
	Codling, E.~A. and Bode, N. W.~F. (2016).
	\newblock {Balancing direct and indirect sources of navigational information in
		a leaderless model of collective animal movement}.
	\newblock {\em Journal of Theoretical Biology}, 394(7 April):32--42.
	
	\bibitem[Codling et~al., 2007]{Codling.ea.2007}
	Codling, E.~A., Pitchford, J.~W., and Simpson, S.~D. (2007).
	\newblock {Group navigation and the ``Many-wrongs pribciple'' in models of
		animal movement }.
	\newblock {\em Ecology}, 88(7):1864--1870.
	
	\bibitem[Dell'Ariccia et~al., 2008]{DellAriccia.ea.2008}
	Dell'Ariccia, G., Dell'Omo, G., Wolfer, D.~P., and Lipp, H.-P. (2008).
	\newblock {Flock flying improves pigeons? homing: GPS track analysis of
		individual flyers versus small groups}.
	\newblock {\em Animal Behavior}, 76(4):1165--1172.
	
	\bibitem[Dodson et~al., 2013]{Dodson.ea.2013}
	Dodson, C.~A., Hore, P.~J., and Wallace, M.~I. (2013).
	\newblock {A radical sense of direction: signalling and mechanism in
		cryptochrome magnetoreception}.
	\newblock {\em Trends in Biochemical Sciences}, 38(9):435--446.
	
	\bibitem[Faria et~al., 2009]{Faria.ea.2009}
	Faria, J.~J., Codling, E.~A., Dyer, J. R.~G., Trillmich, F., and Krause, J.
	(2009).
	\newblock {Navigation in human crowds; testing the many-wrongs principle}.
	\newblock {\em Animal Behaviour}, 78:587--591.
	
	\bibitem[Flack et~al., 2015]{Flack.ea.2015}
	Flack, A., Biro, D., Guilford, T., and Freeman, R. (2015).
	\newblock {Modelling group navigation: transitive social structures improve
		navigational performance}.
	\newblock {\em J. R. Soc. Interface}, 12(20150213):1--8.
	
	\bibitem[Galler et~al., 1972]{Galler.1972}
	Galler, S.~R., Schmidt-Koenig, K., Jacobs, G.~J., and Belleville, R.~E.,
	editors (1972).
	\newblock {\em Animal Orientation and Navigation}.
	\newblock National Aeronautics and Space Administration, U.S.A., Washington.
	
	\bibitem[G\"{o}k\c{c}e and \c{S}ahin, 2010]{Gokce.ea.2010}
	G\"{o}k\c{c}e, F. and \c{S}ahin, E. (2010).
	\newblock {The pros and cons of flocking in the long-range ``migration'' of
		mobile robot swarms}.
	\newblock {\em Theoretical Computer Science}, 411:2140--2154.
	
	\bibitem[Gould, 1982]{Gould.1982}
	Gould, J.~L. (1982).
	\newblock The map sense of pigeons.
	\newblock {\em Nature}, 296(5854):205--211.
	
	\bibitem[Gould, 2011]{Gould.2011}
	Gould, J.~L. (2011).
	\newblock Animal navigation: Longitude at last.
	\newblock {\em Current Biology}, 21(6):R225--R227.
	
	\bibitem[Keeton, 1970]{Keeton.1970}
	Keeton, W.~T. (1970).
	\newblock {Comparative orientational and homing performances of single pigeons
		and small flocks}.
	\newblock {\em Auk}, 87:797--799.
	
	\bibitem[Kirschvink and Walker, 1985]{Kirschvink.and.Walker.1985}
	Kirschvink, J.~L. and Walker, M.~M. (1985).
	\newblock {Particle-size considerations for magnetite-based magnetoreceptors}.
	\newblock In Kirschvink, J.~L., Jones, D.~S., and MacFadden, B.~J., editors,
	{\em {Magnetite Biomineralization and Magnetoreception by Living Organisms: A
			New Biomagnetism}}, pages 243--254. Plenum Press, New York.
	
	\bibitem[Kishkinev et~al., 2013]{Kishkinev.ea.2013}
	Kishkinev, D., Chernetsov, N., Heyers, D., and Mouritsen, H. (2013).
	\newblock Migratory reed warblers need intact trigeminal nerves to correct for
	a 1,000 km eastward displacement.
	\newblock {\em PLoS One}, 8(6):e65847.
	
	\bibitem[Kishkinev and Chernetsov, 2015]{Kishkinev.ea.2015}
	Kishkinev, D.~A. and Chernetsov, N.~S. (2015).
	\newblock Magnetoreception systems in birds: A review of current research.
	\newblock {\em Biology Bulletin Reviews}, 5(1):46--62.
	
	\bibitem[Nagy et~al., 2010]{Nagy.ea.2010}
	Nagy, M., \'{A}kos, Z., Biro, D., and Vicsek, T. (2010).
	\newblock {Hierarchical group dynamics in pigeon flocks}.
	\newblock {\em Nature}, 464(7290):890--894.
	
	\bibitem[Ritz et~al., 2009]{Ritz.ea.2009}
	Ritz, T., Wiltschko, R., Hore, P.~J., Rodgers, C.~T., Staputt, K., Thalau, P.,
	Timmel, C.~R., and Wiltschko, W. (2009).
	\newblock Magnetic compass of birds is based on a molecule with optimal
	directional sensitivity.
	\newblock {\em Biophysical Journal}, 96:3451--3457.
	
	\bibitem[Rodrigo, 2002]{Rodrigo.2002}
	Rodrigo, T. (2002).
	\newblock {Navigational strategies and models}.
	\newblock {\em Psicol\'{o}gica}, 23(1):3--32.
	
	\bibitem[Shaw et~al., 2015]{Shaw.ea.2015}
	Shaw, J., Boyd, A., House, M., Woodward, R., Mathes, F., Cowin, G., Saunders,
	M., and Baer, B. (2015).
	\newblock {Magnetic particle-mediated magnetoreception}.
	\newblock {\em J. R. Soc. Interface}, 12(20150499).
	
	\bibitem[Simons, 2004]{Simons.2004}
	Simons, A.~M. (2004).
	\newblock {Many wrongs: the advantage of group navigation}.
	\newblock {\em Trends in Ecology and Evolution}, 19(9):453--455.
	
	\bibitem[Tamm, 1980]{Tamm.1980}
	Tamm, S. (1980).
	\newblock {Bird orientation: single homing pigeons compared with small flocks}.
	\newblock {\em Behav. Ecol. Sociobiol.}, 7(4):319--322.
	
	\bibitem[Vicsek et~al., 1995]{Vicsek.ea.1995}
	Vicsek, T., Czirok, A., Ben-Jacob, E., Cohen, I., and Shochet, O. (1995).
	\newblock Novel type of phase transition in a system of self-driven particles.
	\newblock {\em Physical Review Letters}, 75(6):1226--1229.
	
	\bibitem[Wallraff, 1978]{Walraff.1978}
	Wallraff, H.~G. (1978).
	\newblock {Social interrelations involved in migratory orientation of birds:
		possible contribution of field studies}.
	\newblock {\em Oikos}, 30:401--404.
	
	\bibitem[Wiltschko and Wiltschko, 1995]{Wiltschko.and.Wiltschko.1995}
	Wiltschko, R. and Wiltschko, W. (1995).
	\newblock {\em Magnetic Orientation in Animals}.
	\newblock Springer, Berlin.
	
	\bibitem[Wiltschko and Wiltschko, 1972]{Wiltschko.ea.1972}
	Wiltschko, W. and Wiltschko, R. (1972).
	\newblock Magnetic compass of european robins.
	\newblock {\em Science}, 176(4030):62--64.
	
	\bibitem[Wystrach et~al., 2016]{Wystrach.ea.2016}
	Wystrach, A., Dewar, A., Philippides, A., and Graham, P. (2016).
	\newblock {How do field of view and resolution affect the information content
		of panoramic scenes for visual navigation? A computational investigation}.
	\newblock {\em J. Comp. Physiol. A}, 202(2):87--95.
	
\end{thebibliography}

\end{document}